\documentclass[lettersize,journal]{IEEEtran}
\usepackage{cite}
\usepackage{amsmath,amssymb,amsfonts}
\usepackage{algorithmic}
\usepackage{graphicx}
\usepackage{textcomp}

\begin{document}
\title{A Global Asymptotic Convergent Observer for SLAM}
\author{Seyed Hamed Hashemi, Jouni Mattila
\thanks{Manuscript received XXX; revised XXX; accepted XXX. Date of publication XXX; date of current version XXX. This work was supported in part by the XXX. Recommended by Associate Editor XXX (Corresponding author: Jouni Mattila.)}
\thanks{The authors are with the Faculty of Engineering and Natural Sciences, Unit of Automation Technology and Mechanical Engineering, Tampere University, Tampere, Finland (e-mail: hamed.hashemi@tuni.fi, jouni.mattila@tuni.fi).}}

\maketitle

\begin{abstract}
This paper examines the global convergence problem of SLAM algorithms, an issue that faces topological obstructions. This is because the state-space of attitude dynamics is defined on a non-contractible manifold: the special orthogonal group of order three $SO(3)$. Therefore, this paper presents a novel, gradient-based hybrid observer to overcome these topological obstacles. The Lyapunov stability theorem is used to prove the globally asymptotic convergence of the proposed algorithm. Finally, comparative analyses of two simulations were conducted to evaluate the performance of the proposed scheme and to demonstrate the superiority of the proposed hybrid observer to a smooth observer.  
\end{abstract}


\section{Introduction}
\IEEEPARstart{S}{imultaneous} Localization and Mapping (SLAM) is a well-known highly nonlinear problem which many previous studies have examined \cite{makinen2019application}. This estimation problem has an extensive variety of applications, ranging from unmanned aerial vehicles (UAV) to underwater robotics. Likewise, the co-dependence of environmental mapping and pose estimation makes the problem of significant theoretical interest. In the SLAM problem, a mobile robot tries to construct a map of an environment while simultaneously estimating its pose (i.e., attitude and position) \cite{bailey2006simultaneous}. Different types of estimation techniques have been applied to the SLAM problem, including Kalman-type filters \cite{castellanos2004limits} and geometric nonlinear observers \cite{mahony2017geometric}.

As mentioned above, the Kalman filter and its variants are the estimation algorithms that have most frequently been employed to solve SLAM \cite{huang2007convergence}. Nevertheless, Kalman-type filters suffer from some serious shortcomings; for example, their performance depends on the prior information regarding noise statistics and initial values; consistency is also an issue \cite{hashemi2019doppler1}. Several previous studies have addressed these limitations \cite{hashemi2019doppler2}. For example, \cite{huang2013quadratic} introduced a new unscented Kalman-type filter (UKF), called observability-constrained (OC)-UKF, to address two key restrictions of UKF: the computational cost in high-dimension systems and the inconsistency problem. A right invariant extended Kalman ﬁlter (RI-EKF) based on a new Lie group structure has also been presented by \cite{song2021right} to solve the inconsistency issue. Masreliez-Martin UKF (MMUKF) has been presented in \cite{tang2020robot} to resolve the problems related to stability and tracking accuracy. In this strategy, an adaptive factor was included to calculate the process noise covariance matrix, and a dynamic model of the robot was utilized to predict the locations of the robot and of landmarks. The inconsistency of EKF-SLAM has also been investigated by \cite{huang2008analysis}; here, the filter Jacobians are determined utilizing the first-ever accessible estimates for each state to preserve the dimensions of the observable subspace. \cite{he2018slam} used a combination of EKF and a particle filter to address the SLAM problem. In this method, the particle filter determines the position of the mobile robot, and the EKF estimates the position of the environment. The performance of the UKF-SLAM was further developed by \cite{bahraini2020efficiency}, who rendered an adaptive random search maximization scheme to adapt the scaling parameter. To further improve the performance of the standard UKF-SLAM and reduce its dependency on prior knowledge, a robust SLAM has also been developed based on $H_{\infty}$ square root UKF \cite{havangi2016robust}.

One recently adopted technique for solving the SLAM problem is the use of geometric nonlinear observers. In these strategies, observers are directly designed in matrix Lie groups, including $SE_{1+n}(3)$ and $SLAM_{n}(3)$. For instance, \cite{zlotnik2018gradient} used a gradient-based observer in the underlying Lie group; the innovation term was derived from the descent direction of an error function. Utilizing group speed and landmark measurements, a geometric nonlinear observer has been introduced that evolves directly on the matrix Lie group $SE_{1+n}(3)$ \cite{wang2018geometric}. Furthermore, \cite{hashim2020landmark} developed a geometric nonlinear observer directly on the manifold of the Lie group $SLAM_{n}(3)$. This proposed observer could guarantee predefined performance parameters and remove the unspecified bias in velocity measurements by using measurements of inertial measurement unit (IMU), group velocity, and landmarks. In a continuation of previous work, the authors have developed the observer by diminishing the boundaries of the error function to ensure faster convergence on the origin \cite{hashim2020guaranteed}. The SLAM manifold has been introduced to develop the matrix Lie group $SLAM_{n}(3)$ for the SLAM problem in \cite{mahony2021homogeneous}. Consequently, a global asymptotic stable observer has also been derived on the suggested manifold to solve SLAM in dynamic environments.

Alongside the SLAM problem, Visual SLAM (VSLAM) has also received significant attention. VSLAM is a specific case of SLAM in which a monocular camera provides the measurements. Van Goor \textit{et al.} proposed a new Lie group called $VSLAM_{n}(3)$ and derived an almost globally asymptotically stable observer on $VSLAM_{n}(3)$ \cite{van2019geometric}. The introduced observer utilized decoupled gain matrices for each landmark and employed a new cost function to calculate innovations in the robot's pose. In addition, \cite{van2020observer} continues the authors’ prior work; here, a gradient-based observer with almost global stability has been designed in the $VSLAM_{n}(3)$ Lie group. Van Goor's \textit{et al.} \cite{van2019geometric} work has been further developed in \cite{van2021constructive} with the introduction of equivariant group actions. The suggested nonlinear equivariant observer's almost semi-global convergence is its most important feature.

Although the observers described above have a number of advantages, they also share a significant shortcoming. To the best of the authors’ knowledge, most state-of-the-art observers ensure almost global stability \cite{wang2018hybrid}. This is because the special orthogonal group of order three $SO(3)$ is a non-contractible manifold \cite{hashemi2021observer}. Hence, there exists a set with 
Lebesgue measure zero from which the estimation error cannot converge to zero. Hybrid systems have therefore been used to overcome this topological obstruction and to derive observers with global stability on $SO(3)$ \cite{hashemi2022observer}, $SE(3)$ \cite{wang2017globally}, and $SE_2(3)$ \cite{wang2018globally}. For example, two hybrid observers were introduced in \cite{wang2020hybrid}; the first observer uses fixed gains, while the second uses variable gains by solving a continuous Riccati equation. Wang \textit{et al.} \cite{wang2020nonlinear} expands on the authors’ previous work;
here, the same strategy has been used to develop two hybrid observers. In contrast to previous observers, these observers do not need information about the gravity vector and can overcome the difficulty in estimation under intermittent landmark measurements.

In light of the shortcomings of previous solutions, the present paper addresses the problem of designing an observer with global stability for SLAM. Here, a new, gradient-based hybrid observer is introduced on the $SE_{1+n}(3)$ manifold to solve the SLAM problem. The present article is divided into five sections, including the introduction. Section 2 provides the preliminary mathematical notation, SLAM kinematics and measurements equations, and the basic background on hybrid systems. The proposed hybrid observer is described in section 3. Section 4 illustrates the results of simulations in which the proposed observer is compared to a smooth observer. Finally, section 6 summarizes the paper and provides some concluding remarks.

\section{Preliminaries }
\subsection{Notation}
This paper denotes the sets of real, nonnegative real, and natural numbers by $\mathbb{R}$, $\mathbb{R}_{\ge0}$, and $\mathbb{N}$, respectively. ${{\mathbb{R}}^{n}}$ represents $n$-dimensional Euclidean space, where $\{e_i\}_{1\le i \le n} \subset {{\mathbb{R}}^{n}}$ is the canonical basis of  ${{\mathbb{R}}^{n}}$. $\left\| x \right\|=\sqrt{\left\langle x,x \right\rangle }$ denotes the two-norm of a vector where $\left\langle x,y \right\rangle :={{x}^{T}}y$ is the inner products of vectors $x,y\in {{\mathbb{R}}^{n}}$ and ${{\left\| x \right\|}_{\mathcal{A}}}:={{\min }_{y\in \mathcal{A}}}\left\| x-y \right\|$. The trace, determinant, and transpose of a matrix $A \in \mathbb{R}^{n \times n}$ are denoted by $\text{tr}(A)$, $\text{det}(A)$, and $A^T$, respectively. Moreover, $\left\| A \right\|_F=\sqrt{\left\langle A,A \right\rangle}$ is the Frobenius norm of $A$, where $\left\langle A,B \right\rangle := \text{tr}(A^{T}B)=(\text{vec}A)^{T}(\text{vec}B)$, and $\text{vec}A=[Ae_{1} \ldots Ae_{n}]^T$ is the vectorization of $A$. The attitude of a rigid body is denoted by $R \in SO(3)$, where $SO(3):=\{R \in \mathbb{R}^{3\times3}:R^TR=RR^T=I, \text{det}(R)=1 \}$ is the special orthogonal group of order three, and $\mathfrak{so}(3)=\{A \in \mathbb{R}^{3 \times 3}: A^T=-A \}$ is the Lie algebra of $SO(3)$. In this paper, $SE_{1+n}(3):=\{ \mathcal{X}= \Psi(R,p,\eta):R \in SO(3), p \in \mathbb{R}^3, \eta \in \mathbb{R}^{3\times n}\}$ represents the matrix Lie group. Throughout this paper, the following identities are used frequently.

\begin{equation}\label{eq1}
	\begin{split}
		&\Gamma (y)=\left[ \begin{matrix}
			0 & -{{y}_{3}} & {{y}_{2}}  \\
			{{y}_{3}} & 0 & -{{y}_{1}}  \\
			-{{y}_{2}} & {{y}_{1}} & 0  \\
		\end{matrix} \right], \varphi(A)=\frac{1}{2} \left[ \begin{matrix}
			A_{(3,2)}-A_{(2,3)} \\
			A_{(1,3)}-A_{(3,1)} \\
			A_{(2,1)}-A_{(1,2)} \\
		\end{matrix} \right],\\
		&\Psi(R,p,\eta)= \left[\begin{array}{c c|c} 
			R & p & \eta \\
			0_{1\times3} & 1 & 0_{1\times n} \\
			\hline 
			0_{n\times3} & 0_{n\times1} & I_{n \times n} 
		\end{array}\right], \\
		&\Upsilon(B)=\Upsilon(\left[ \begin{matrix}
			B_1 & B_2 \\
			B_3^T & B_4 \\
		\end{matrix} \right])= \left[ \begin{matrix}
			\frac{1}{2}(B_1-B_1^T) & B_2 \\
			0_{n+1 \times 3} & 0_{n+1 \times n+1} \\
		\end{matrix} \right],\\
		&y \in \mathbb{R}^3, A, B_1 \in \mathbb{R}^{3 \times3} ,B_2, B_3 \in \mathbb{R}^{3 \times n+1}, B_4 \in \mathbb{R}^{n+1 \times n+1} \\
	\end{split}
\end{equation}

The inverse of $\mathcal{X}$ is determined as $ {\mathcal{X}}^{-1} = \Psi (R^T,-R^Tp,-R^T \eta)$, and the Lie algebra associated with the $SE_{1+n}(3)$ is given by

\begin{equation*}\label{eq2}
	\begin{split}
		&\mathfrak{se}_{1+n}(3):=\{\mathcal{V}(\omega,v,\xi)= \left[\begin{array}{c c|c} 
			\Gamma(\omega) & v & \xi \\
			\hline 
			0_{n+1\times3} & 0_{n+1\times1} & 0_{n+1 \times n} 
		\end{array}\right]:\\
		& \omega, v \in \mathbb{R}^3, \xi \in \mathbb{R}^{3\times n} \}.\\
	\end{split}
\end{equation*}

The gradient of a differentiable smooth function $m:SE_{1+n}(3)\to \mathbb{R}$ is denoted by $\nabla_{\mathcal{X}}m \in T_{\mathcal{X}}SE_{1+n}(3)$, where $T_{\mathcal{X}}SE_{1+n}(3):= \{\mathcal{X} \mathcal{V}: \mathcal{X} \in SE_{1+n}(3), \ \mathcal{V} \in \mathfrak{se}_{1+n}(3)\}$ is the tangent space of the $SE_{1+n}(3)$. Accordingly, $\nabla_{\mathcal{X}}m$ is calculated using the following equation. 

\begin{equation}\label{eq3}
	dm.\mathcal{X}\mathcal{V}= \left\langle \nabla_{\mathcal{X}}m,\mathcal{X}\mathcal{V} \right\rangle_\mathcal{X}=\left\langle \mathcal{X}^{-1} \nabla_{\mathcal{X}}m,\mathcal{V} \right\rangle
\end{equation}

Where $dm$ is the differential of $m$ and $\left\langle.,.\right\rangle_\mathcal{X}$ is a Riemannian metric on $SE_{1+n}(3)$ such that

\begin{equation*}\label{eq4}
	\left\langle \mathcal{X}\mathcal{V}_1,\mathcal{X}\mathcal{V}_2 \right\rangle_\mathcal{X} = \left\langle \mathcal{V}_1,\mathcal{V}_2 \right\rangle.
\end{equation*}

The adjoint map $Ad_{\mathcal{X}}:SE_{1+n}(3) \times \mathfrak{se}_{1+n}(3) \rightarrow \mathfrak{se}_{1+n}(3)$ is defined as $Ad_{\mathcal{X}}\mathcal{V}:=\mathcal{X}\mathcal{V}\mathcal{X}^{-1}$; this takes a tangent vector of one element and transforms it to a tangent vector of another element. The Rodrigues formula $\Re :\mathbb{R}\times {\mathbb{S}^{2}}\to SO(3)$ parametrizes a rotation matrix $R \in SO(3)$ using a specific angle $\theta \in \mathbb{R}$ around a fixed axis $y\in {\mathbb{S}^{2}}$; this formula is expressed as follows.

\begin{equation}\label{eq5}
	\Re (\theta ,y)=I+\sin (\theta )\Gamma (y)+(1-\cos (\theta )){{\Gamma }^{2}}(y)=\exp(\theta \Gamma(y))
\end{equation}

Where $\mathbb{S}^2:=\{y\in \mathbb{R}^3: \|y\|=1 \} $ is a unit two-dimensional sphere.

\subsection{SLAM Kinematics}
The kinematic equations define the motion of a rigid body, and a family of $n$ landmarks are given as follows.

\begin{equation}\label{eq6}
	\dot{R}=R\Gamma (\omega )
\end{equation}
\begin{equation}\label{eq7}
	\dot{p}=Rv
\end{equation}
\begin{equation}\label{eq8}
	\quad \quad	\quad \dot{\eta_i}=R\xi_i, \quad i=1,\dots,n
\end{equation}

Here, $\omega \in \mathbb{R}^3$ and $v \in \mathbb{R}^3$ are the angular rate and linear velocity of the rigid body expressed in the body-fixed frame $\mathcal{B}$, respectively. $\xi_i \in \mathbb{R}^3$ is the linear speed of the $i$-th landmark expressed in $\mathcal{B}$. The kinematic equations (\ref{eq6})- (\ref{eq8}) can be rephrased using the following compact form.

\begin{equation}\label{eq9}
	\dot{\mathcal{X}}=\mathcal{X}\mathcal{V}
\end{equation}

In this paper, it is assumed that the landmarks are stationary (i.e., $\xi_i=0$) and that the linear and angular velocities of the rigid body are available for measurement. It is also assumed that the angular and linear velocity measurements include an unknown constant bias, that is

\begin{equation}\label{eq10}
	\begin{split}
		& \mathcal{V}_m = \mathcal{V}+\mathcal{V}_b, \\ 
		& \mathcal{V}_m=\mathcal{V}(\omega_m,v_m,0), \ \mathcal{V}_b=\mathcal{V}(b_{\omega},b_v,0), \ b=[b_{\omega} \ b_v]^T.\\
	\end{split}
\end{equation}

Furthermore, it is also assumed that the robot can perceive both range $\gamma=\|\eta_i-p\|$ and bearing $\jmath=R^T(\eta_i-p)/\gamma$ relative to landmarks. Accordingly, the following compact equation is the result of a combination of the range and bearing measurements.

\begin{equation}\label{eq11}
	\begin{split}
		& \beta_i=\mathcal{X}^{-1}r_i=\left[\begin{array}{c} 
			R^T(\eta_i-p) \\
			1 \\
			-e_i 
		\end{array}\right], \quad i=1,\dots,n \\
		& r_i=\left[\begin{array}{c} 
			0_{3\times1} \\
			1 \\
			-e_i 
		\end{array}\right] \\
	\end{split}
\end{equation}

\subsection{Hybrid System Frameworks}
The present paper uses the following framework of hybrid systems $\mathcal{H}$ that was first introduced by \cite{4806347}.

\begin{equation}\label{eq12}
	\mathcal{H}:\left\{ \begin{matrix}
		\dot{x}=f(x,u), & (x,u)\in C \\
		{{x}^{+}}=g(x,u), & (x,u)\in D \\
	\end{matrix} \right.
\end{equation} 

In this framework, $f:{{\mathbb{R}}^{n}}\times {{\mathbb{R}}^{m}}\to {{\mathbb{R}}^{n}}$ is the flow map that defines the continuous dynamics of $\mathcal{H}$, and $g:{{\mathbb{R}}^{n}}\times {{\mathbb{R}}^{m}}\to {{\mathbb{R}}^{n}}$ is the jump map that specifies the behavior of $\mathcal{H}$ during jumps. The flow set $C\subset {{\mathbb{R}}^{n}}\times {{\mathbb{R}}^{m}}$ indicates where continuous evolution is allowed to flow, and the jump set $D\subset {{\mathbb{R}}^{n}}\times {{\mathbb{R}}^{m}}$ demonstrates where the system is permitted to jump. The subset $E \subset {{\mathbb{R}}_{\ge 0}}\times \mathbb{N}$ is called a hybrid time domain if $E=\bigcup\limits_{i=1}^{I}{(\left[ {{t}_{i}},{{t}_{i+1}} \right],i)}$ for finite sequences of times $0={{t}_{0}}\le {{t}_{1}}\cdots \le {{t}_{I+1}}$. A hybrid arc consists of a hybrid time domain $\text{dom }x$ and a function $x:\text{dom }x\to {{\mathbb{R}}^{n}}$, which is also called a solution to $\mathcal{H}$. 

\textbf{Lemma} \cite{teel2012lyapunov}: The closed set $\mathcal{A} \subset \mathbb{R}^n$ is locally and exponentially stable for $\mathcal{H}$, if $(\alpha_1 > \alpha_2, s_1, s_2, n)\in \mathbb{R}_{\ge0}$ exists and there is a continuously differentiable function $V:\text{dom }V\to {{\mathbb{R}}}$ on an open set containing the
closure of $C$ that satisfies the following equation.

\begin{equation}\label{eq13}
	\begin{split}
		& \alpha_2{\left\| x \right\|}_{\mathcal{A}}^n \le V(x) \le \alpha_1{\left\| x \right\|}_{\mathcal{A}}^n, \\
		& \forall x \in (C \cup D \cup g(D)) \cap (\mathcal{A}+s_1\mathbb{B})\\
		&\left\langle \nabla V(x),f \right\rangle \le -s_2V(x), \quad \forall x \in C \cap (\mathcal{A}+s_1\mathbb{B})\\
		& V(g) \le \exp(-s_2)V(x), \quad \forall x \in D \cap (\mathcal{A}+s_1\mathbb{B}).\\
	\end{split}
\end{equation}

where $\mathbb{B} :=\{x\in \mathbb{R}^n: \|x\| \le1 \}$ is the closed unit ball. The set $\mathcal{A}$ is said to be globally exponentially stable if $s_1=\infty$, and $\mathcal{A}$ is said to be globally asymptotically stable if $s_1=\infty, s_2=0$.

\section{Proposed Hybrid Observer}
This section describes the proposed observer. As mentioned above, the two main techniques that have been utilized to solve the SLAM problem are geometric nonlinear observers and Kalman-type filters. The drawbacks of these methods have also been discussed in the previous section. Most state-of-the-art observers are almost globally stable due to the non-contractibility of the state-space of attitude kinematics (i.e., $SO(3)$). Consequently, hybrid systems have been used to tackle this topological challenge and to obtain globally stable results \cite{hashemi2021global}. Therefore, the present paper builds on the observer developed by \cite{wang2018geometric} and describes a hybrid observer for solving the SLAM problem. Consider the following smooth real-value function: $\mathcal{U}: SE_{1+n}(3) \rightarrow \mathbb{R}$.

\begin{equation}\label{eq14}
	\mathcal{U}(\mathcal{X})=\frac{1}{2}\text{tr}((I-\mathcal{X})A(I-\mathcal{X})^T)
\end{equation}

Where $A:=\sum_{i=1}^{n} k_ir_i{r_i}^T$ and $k_i \in \mathbb{R}_{\ge0}$ are positive constants. Utilizing the Riemannian metric on $SE_{1+n}(3)$ and the identities provided in the Appendix \ref{APP1}, one can show that

\begin{equation}\label{eq15}
	\begin{split}
		& d\mathcal{U}.\mathcal{X}\mathcal{V}=\left\langle \mathcal{X}^{-1} \nabla_{\mathcal{X}}\mathcal{U},\mathcal{V} \right\rangle \Rightarrow\\
		& d\mathcal{U}.\mathcal{X}\mathcal{V}=\text{tr}(-A(I-\mathcal{X})^T\mathcal{X}\mathcal{V}) \\
		& =\left\langle \Upsilon(\mathcal{X}^{-1}(\mathcal{X}-I)A),\mathcal{V} \right\rangle \\
		& =\left\langle \Upsilon((I-\mathcal{X}^{-1})A),\mathcal{V} \right\rangle. \\
	\end{split}
\end{equation}

Therefore, the gradient of $\mathcal{U}$ with respect to $\mathcal{X}$ is calculated with the following equation. 

\begin{equation}\label{eq16}
	\nabla_{\mathcal{X}}(\mathcal{U})=\mathcal{X}\Upsilon((I-\mathcal{X}^{-1})A)
\end{equation}

Throughout this paper, $\hat{\mathcal{X}}$ denotes the estimated value of the state $\mathcal{X}$. Therefore, $\tilde{\mathcal{X}}=\hat{\mathcal{X}}\mathcal{X}^{-1}$ is the estimation error with $\tilde{R}=R\hat{R}^T$, $\tilde{p}=p-\tilde{R}\hat{p}$, and $\tilde{\eta}=\eta-\tilde{R}\hat{\eta}$. Hence, the following identities can be easily calculated using (\ref{eq9}) and (\ref{eq1}). (For details, see \cite{wang2018geometric}.)

\begin{equation}\label{eq17}
	\begin{split}
		& \sum_{i=1}^{n} k_i\|r_i-\hat{\mathcal{X}}\beta_i\|^2=\text{tr}((I-\tilde{\mathcal{X}})A(I-\tilde{\mathcal{X}})^T), \\
		& \Upsilon(\sum_{i=1}^{n} k_i(r_i-\hat{\mathcal{X}}\beta_i)r_i^T)=\Upsilon((I-\tilde{\mathcal{X}}^{-1})A) \\
	\end{split}
\end{equation}

The dynamics of the proposed hybrid observer are defined as follows.

\begin{equation}\label{eq18}
	\begin{split}
		&\begin{cases}
			\dot{\hat{\mathcal{X}}}=\hat{\mathcal{X}}(\mathcal{V}_m-\mathcal{V}_{\hat{b}}-\Delta), \\
			\dot{\mathcal{V}}_{\hat{b}}=-\Upsilon(\hat{\mathcal{X}}^T\sum_{i=1}^{n} k_i(r_i-\hat{\mathcal{X}}\beta_i)r_i^T\hat{\mathcal{X}}^{-T})K, & (\hat{\mathcal{X}},\hat{b}) \in C\\
			\dot{q}=0, \\
		\end{cases}\\ 
		&\begin{cases}
			\hat{\mathcal{X}}^+=\mathcal{X}_q, \\
			\mathcal{V}_{\hat{b}}^+=\mathcal{V}_{\hat{b}}, & (\hat{\mathcal{X}},\hat{b}) \in D \\
			q^+=\underset{q \in \mathcal{Q}}{\arg\min} \ \mathcal{U}(\tilde{\mathcal{X}_q}),
		\end{cases}\\
		& C:=\{\mathcal{U}(\tilde{\mathcal{X}})-\min_{\tilde{\mathcal{X}_q}\in \mathcal{Q}}  \mathcal{U}(\tilde{\mathcal{X}_q})\le \delta\}, \\
		& D:=\{\mathcal{U}(\tilde{\mathcal{X}})-\min_{\tilde{\mathcal{X}_q}\in \mathcal{Q}} \mathcal{U}(\tilde{\mathcal{X}_q})\ge \delta\}, \\
		&\mathcal{X}_q=\Psi(\Re(0.2q\theta,\ell)'\hat{R},\Re(0.2q\theta,\ell)\hat{p},2q\hat{\eta}), \quad q \in \mathbb{N}\\
		&\Delta=-Ad_{\hat{\mathcal{X}}^{-1}}\Upsilon(\sum_{i=1}^{n} k_i(r_i-\hat{\mathcal{X}}\beta_i)r_i^T),\\
		& \hat{b} = [\varphi(\mathcal{V}_{\hat{b}}(1:3,1:3)) \ \mathcal{V}_{\hat{b}}(1:3,4)]^T \\
	\end{split}
\end{equation}

In equation (\ref{eq18}), $\theta,\delta \in \mathbb{R}_{>0}$ are arbitrary constants, $\ell \in \mathbb{S}^2$ is an arbitrary fixed vector, $\mathcal{Q}=\{\mathcal{X}_q \in SE_{1+n}(3): q \in \mathbb{N} \}$ is a compact set, $K:=k_o I_{n \times n}$ with $0<k_o<1 \in \mathbb{R}$ is the observer gain, and $\tilde{\mathcal{X}_q}=\mathcal{X}_q\mathcal{X}^{-1}$.

\textbf{Theorem}: Consider the proposed hybrid observer (\ref{eq18}) with any $\theta \in \mathbb{R}_{>0}$ and $\ell \in \mathbb{S}^2$ for the SLAM kinematics (\ref{eq9}). The state estimation error $\tilde{\mathcal{X}}$ and bias estimation error $\mathcal{V}_{\tilde{b}}=\mathcal{V}_b-\mathcal{V}_{\hat{b}}$ converge to $I_{n \times n}$ and 0, respectively; therefore, the following set is globally asymptotically stable.

\begin{equation}\label{eq19}
	\mathcal{A}:=\{ (\tilde{\mathcal{X}},\mathcal{V}_{\tilde{b}}) \in SE_{1+n}(3) \times \mathfrak{se}_{1+n}(3): \tilde{\mathcal{X}}=I, \mathcal{V}_{\tilde{b}}=0\}
\end{equation}

\textbf{Proof}: According to \textbf{Lemma}, \textbf{Theorem} is proven in two steps. \\
\textbf{Step 1}: This step proves the second condition of (\ref{eq13}) with $s_2=0$. Utilizing the facts $\dot{\hat{\mathcal{X}}}^{-1}=-\hat{\mathcal{X}}^{-1} \dot{\hat{\mathcal{X}}} \hat{\mathcal{X}}^{-1}$, and $\dot{\mathcal{V}}=\dot{\mathcal{V}}_m=0$, one has

\begin{equation}\label{eq20}
	\begin{split}
		&\dot{\tilde{\mathcal{X}}} = (Ad_{\hat{\mathcal{X}}}(\Delta-\mathcal{V}_{\tilde{b}}))\tilde{\mathcal{X}} \\
		&\dot{\mathcal{V}}_{\tilde{b}}=-\dot{\mathcal{V}}_{\hat{b}}. \\
	\end{split}
\end{equation}

Hence, the estimation error dynamics can be calculated using the following equation.

\begin{equation}\label{eq21}
	\begin{split}
		&\dot{\tilde{\mathcal{X}}} = (-\Upsilon((I-\tilde{\mathcal{X}}^{-1})A)-Ad_{\hat{\mathcal{X}}}\mathcal{V}_{\tilde{b}})\tilde{\mathcal{X}} \\
		&\dot{\mathcal{V}}_{\tilde{b}}=\Upsilon(\hat{\mathcal{X}}^T(I-\tilde{\mathcal{X}}^{-1})A\hat{\mathcal{X}}^{-T})K \\
	\end{split}
\end{equation}

The Lyapunov function candidate is defined as follows.

\begin{equation}\label{eq22}
	V(\tilde{\mathcal{X}},\mathcal{V}_{\tilde{b}})=\mathcal{U}(\tilde{\mathcal{X}})+\frac{1}{2}\text{tr}(\mathcal{V}_{\tilde{b}}\mathcal{V}_{\tilde{b}}^T)
\end{equation}

The time derivative of $V$ is calculated as follows.

\begin{equation}\label{eq23}
	\begin{split}
		&\dot{V}=\left\langle \Upsilon((I-\tilde{\mathcal{X}}^{-1})A),(-\Upsilon((I-\tilde{\mathcal{X}}^{-1})A)-Ad_{\hat{\mathcal{X}}} \mathcal{V}_{\tilde{b}} ) \right\rangle \\
		&+\left\langle \Upsilon(\hat{\mathcal{X}}^T(I-\tilde{\mathcal{X}}^{-1})A\hat{\mathcal{X}}^{-T})K,\mathcal{V}_{\tilde{b}} \right\rangle\\
		&= -\left\langle \Upsilon((I-\tilde{\mathcal{X}}^{-1})A),\Upsilon((I-\tilde{\mathcal{X}}^{-1})A) \right\rangle \\
		&-\left\langle \hat{\mathcal{X}}^T(I-\tilde{\mathcal{X}}^{-1})A\hat{\mathcal{X}}^{-T},\mathcal{V}_{\tilde{b}} \right\rangle \\
		&+\left\langle \hat{\mathcal{X}}^T(I-\tilde{\mathcal{X}}^{-1})A\hat{\mathcal{X}}^{-T}K,\mathcal{V}_{\tilde{b}} \right\rangle= \\
		&-\| \Upsilon((I-\tilde{\mathcal{X}}^{-1})A) \|_F^2+(k_o-1)\left\langle \hat{\mathcal{X}}^T(I-\tilde{\mathcal{X}}^{-1})A\hat{\mathcal{X}}^{-T},\mathcal{V}_{\tilde{b}} \right\rangle \\
	\end{split}
\end{equation}

After simplifying (\ref{eq23}) and utilizing the Cauchy–Schwarz inequality for matrix \cite{bourin2006matrix}, the resulting equation is as follows.

\begin{equation}\label{eq24}
	\begin{split}
		&\dot{V}\le -\| \Upsilon((I-\tilde{\mathcal{X}}^{-1})A) \|_F^2 \\
		&\quad \quad \quad +(k_o-1)\| \hat{\mathcal{X}}^T(I-\tilde{\mathcal{X}}^{-1})A\hat{\mathcal{X}}^{-T}\|_{F} \| \mathcal{V}_{\tilde{b}} \|_{F} \le 0 \\
	\end{split}
\end{equation}

It can thus be deduced that $\tilde{R}$, $\tilde{p}$, $\tilde{\eta}$, and $\mathcal{V}_{\tilde{b}}$ are globally bounded. This implies that $\ddot{V}$ is also globally bounded. Barbalat's lemma reveals that $\lim_{t \to +\infty} \dot{V}=0$; therefore, $\tilde{\mathcal{X}}=I$ and $\mathcal{V}_{\tilde{b}}=0$. (For details, see \cite{wang2018geometric}.) \\
\textbf{Step 2}: In this step, the last condition of (\ref{eq13}) is proven. Since the switching variable $q$ generates jumps, it is essential to assay the variation in $V(\tilde{\mathcal{X}},\mathcal{V}_{\tilde{b}})$ to ensure that the Lyapunov function is reduced across jumps. The variation in $V$ along jumps is given by the following equation.

\begin{equation}\label{eq25}
	\begin{split}
		&V(\hat{\mathcal{X}}^+,\mathcal{V}_{\hat{b}}^+)-V(\hat{\mathcal{X}},\mathcal{V}_{\hat{b}})=\mathcal{U}(\hat{\mathcal{X}}^+)-\mathcal{U}(\hat{\mathcal{X}}) \\
		&=\mathcal{U}(\mathcal{X}_q)-\mathcal{U}(\hat{\mathcal{X}}). \\
	\end{split}
\end{equation}

Substituting (\ref{eq25}) into (\ref{eq17}) leads to the following question.

\begin{equation}\label{eq26}
	\mathcal{U}(\tilde{\mathcal{X}_q})-\mathcal{U}(\tilde{\mathcal{X}})=\sum_{i=1}^{n} k_i(\|r_i-\mathcal{X}_q \beta_i\|^2-\|r_i-\hat{\mathcal{X}}\beta_i\|^2)
\end{equation}

From (\ref{eq18}), one can show that

\begin{equation}\label{eq27}
	\min_{\tilde{\mathcal{X}_q}\in \mathcal{Q}} \mathcal{U}(\tilde{\mathcal{X}_q})-\mathcal{U}(\tilde{\mathcal{X}}) \le -\delta.
\end{equation}

Finally, it follows from \textbf{Lemma} that the set $\mathcal{A}$ is globally asymptotically stable.

\section{Simulation Results}
This section presents the numerical simulations used to evaluate the performance of the proposed observer. The proposed hybrid observer is contrasted with the smooth observer described in \cite{wang2018geometric}. The stability of the proposed algorithm is verified under various initial conditions. In both experiments, it is assumed that the robot can measure range and bearing to four landmarks located at

\begin{equation*}
	\eta = \begin{bmatrix}
		10 & 0 & -10 & 0\\
		0 & 15 & 0 & -10\\
		0 & 0 & 0 & 0 \\
	\end{bmatrix}.
\end{equation*}

Moreover, range and bearing measurements contain a noise signal consisting of uniform distribution on the interval [0 0.4] and a Gaussian distribution with zero mean and unit variance. The following constant biases corrupt the angular velocity and linear velocity $b_\omega=[-0.02 \ 0.05 \ 0.03]^T$, $b_v=[0.2 \ 0.05 \ 0.1]^T$, respectively.

\subsection{First Experiment}
\begin{figure}
	\centering
	\includegraphics[width=1\linewidth]{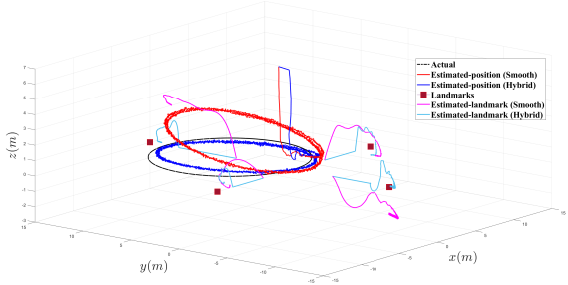}
	\caption{3D trajectories of the observers compared to the actual system evolution in the first experiment.}
	\label{fig1}
\end{figure}

\begin{figure}
	\centering
	\includegraphics[width=1\linewidth]{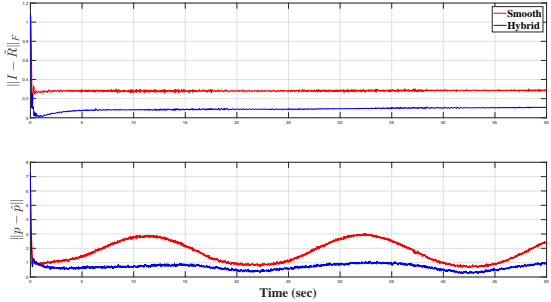}
	\caption{Estimation errors of attitude and position in the first experiment.}
	\label{fig2}
\end{figure}

\begin{figure}
	\centering
	\includegraphics[width=1\linewidth]{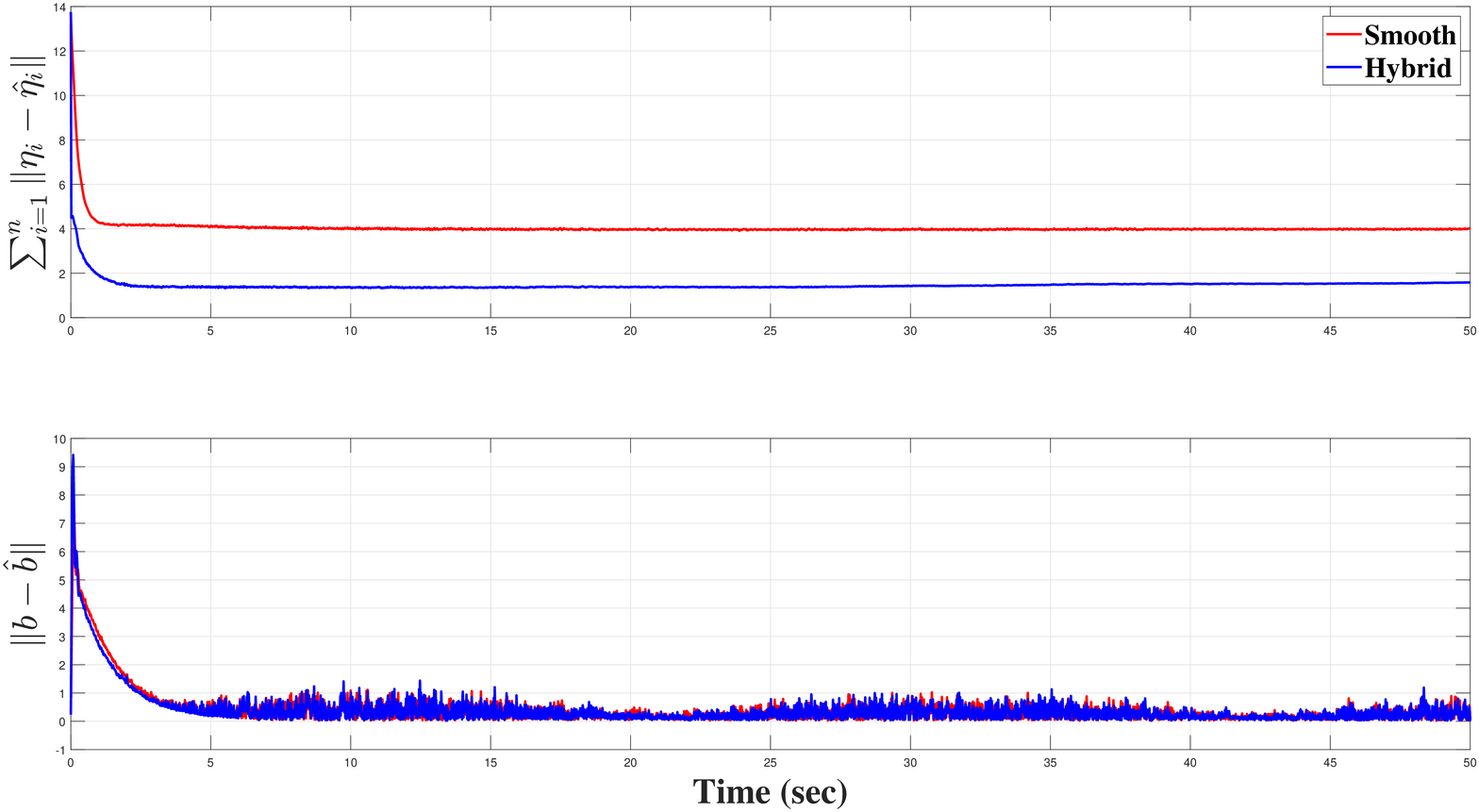}
	\caption{Norms of the velocity bias estimation errors and landmark position estimation errors in the first experiment.}
	\label{fig3}
\end{figure}

\begin{figure}
	\centering
	\includegraphics[width=1\linewidth]{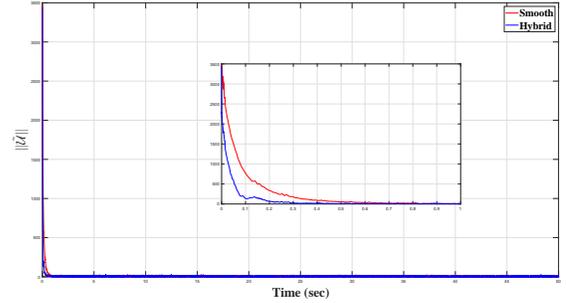}
	\caption{Evolution of the Lyapunov functions versus time in the first experiment.}
	\label{fig4}
\end{figure}

In this experiment, the robot moves in a circular trajectory at a constant altitude. The unbiased measurements of the angular velocity and linear velocity in the body-fixed frame are such that $\omega=[0 \ 0 \ 0.3]^T$ $rad/sec$ and $v=[2 \ 0 \ 0]^T$ $m/sec$. The robot's initial position and attitude were set to $p(0)=[0 \ 0 \ 0]^T$ and $R(0)=\mathfrak{R}(0,e_1)$, respectively. The initial conditions for both observers were set to $\hat{p}(0)=[-2 \ 0 \ 7]^T$, $\hat{R}(0)=\mathfrak{R}(\pi/4,e_1)$, and $\hat{\eta}=0.4*\eta$. Figures (\ref{fig1}-\ref{fig4}) illustrate the results of this experiment. Figure (\ref{fig1}) depicts the estimated paths of the robot's position and the landmarks' positions versus time, as well as the actual robot path and landmark positions. The evolution of the attitude tracking error and position tracking error are shown in Figure (\ref{fig2}). Figure (\ref{fig3}) shows the errors associated with estimates of landmark locations and biases. The evolution of the Lyapunov functions is shown in Figure (\ref{fig4}). These figures demonstrate that the proposed hybrid observer has lower estimation errors than the smooth observer. Furthermore, the convergence rate of the proposed observer is faster than that of the smooth observer.

\subsection{Second Experiment}
\begin{figure}
	\centering
	\includegraphics[width=1\linewidth]{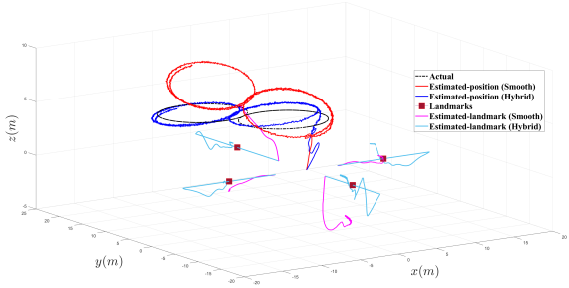}
	\caption{3D trajectories of the observers compared to actual system evolution in the second experiment.}
	\label{fig5}
\end{figure}

\begin{figure}
	\centering
	\includegraphics[width=1\linewidth]{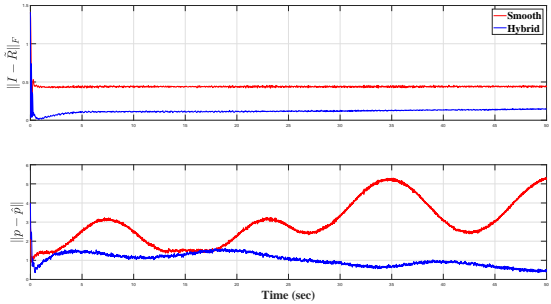}
	\caption{Estimation errors of attitude and position in the second experiment.}
	\label{fig6}
\end{figure}

\begin{figure}
	\centering
	\includegraphics[width=1\linewidth]{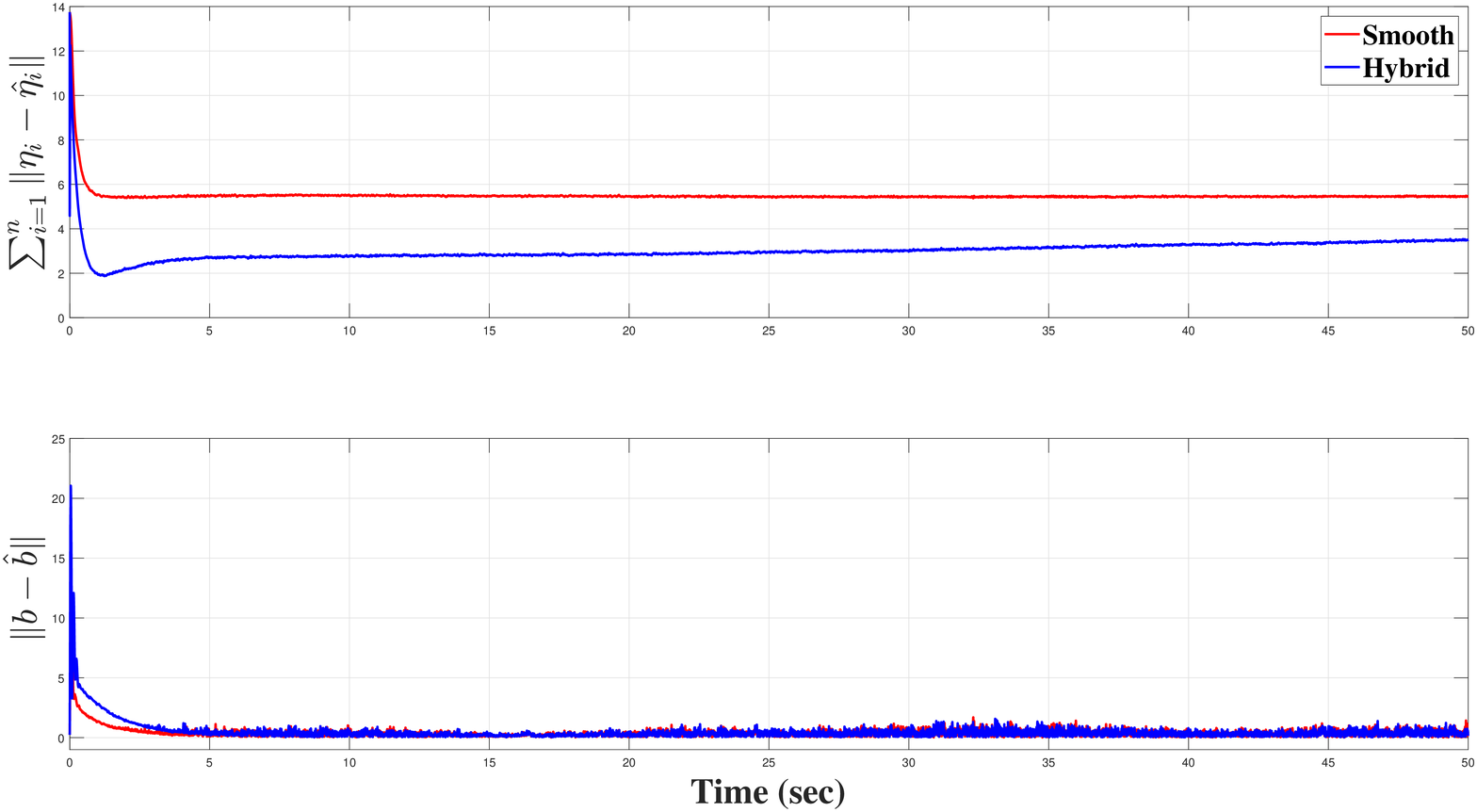}
	\caption{Norms of the velocity bias estimation errors and landmark position estimation errors in the second experiment.}
	\label{fig7}
\end{figure}

\begin{figure}
	\centering
	\includegraphics[width=1\linewidth]{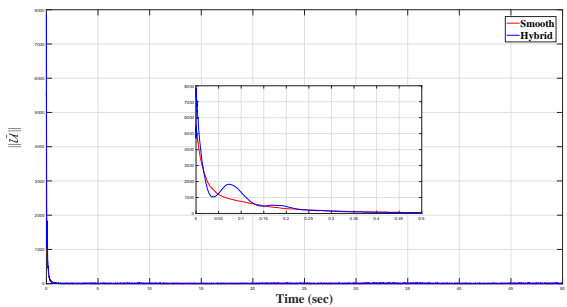}
	\caption{Evolution of the Lyapunov functions versus time in the second experiment.}
	\label{fig8}
\end{figure}

As discussed above, in this experiment, a different simulation scenario is used to further assess the performance of the proposed observer. The robot is simulated to move along a eight-shape trajectory at a fixed height of 4 $m$. The robot's initial attitude and position were set to $R(0)=\mathfrak{R}(0,e_1)$ and $p(0)=[0 \ 0 \ 4]^T$, respectively. In this experiment, angular velocity was assumed to be $\omega=[0 \ 0 \ \pm0.4]^T$ $rad/sec$, and linear velocity was assumed to be $v=[2 \ 0 \ 0]^T$ $m/sec$. Both observers were initialized at $\hat{p}(0)=[0 \ 0 \ 0]^T$, $\hat{R}(0)=\mathfrak{R}(\pi/3,e_1)$, and $\hat{\eta}=0.4*\eta$. The simulation results of the second experiment are shown in Figures (\ref{fig5}-\ref{fig8}). Figure (\ref{fig5}) shows the 3D trajectories of the robot's position tracking and estimated landmark locations, as well as the true robot trajectory and actual landmark locations. The errors in the estimation of attitude, robot path, landmark locations, and biases are illustrated in Figures (\ref{fig6},\ref{fig7}). Figure (\ref{fig8}) shows the evolution of the Lyapunov functions. These figures reveal that the proposed hybrid observer performs better than the smooth observer on both trajectory tracking and reducing the effect of noise.

\section{Conclusion}
The present paper has investigated the problem of global convergence in SLAM observers. Most state-of-the-art SLAM techniques can guarantee almost global convergence due to the non-contractibility of the state-space of attitude. Accordingly, this paper has introduced a gradient-based hybrid observer to overcome topological obstructions and achieve global convergence. The proposed algorithm was demonstrated to be globally asymptotically convergent. Finally, the proposed hybrid observer was compared to a smooth observer, demonstrating the superior performance of the proposed algorithm.

\appendices
\section{Some Useful Identities}\label{APP1}
This paper uses the following identities related to the orthogonal projection and matrix inner product.

\begin{equation*}
\begin{split}
	& \Upsilon(\mathcal{X}B)=\Upsilon(\mathcal{X}^{-T}B), \quad (a) \\
	& \left\langle \mathcal{V},B \right\rangle=\left\langle \mathcal{V},\Upsilon(B) \right\rangle=\left\langle \Upsilon(B),\mathcal{V} \right\rangle, \quad (b)\\
	&\text{tr}(ABCD)=\text{tr}(CDAB)=\text{tr}(DABC), \quad (c)\\
\end{split}
\end{equation*}

\bibliographystyle{ieeetran}
\bibliography{mybibfile}

\end{document}